\documentclass[pra,singlecolumn,nofootinbib,preprint,amssymb,longbibliography,amsfonts,amsmath,superscriptaddress,showpacs,hyperref]{revtex4-1}
\usepackage{graphicx}
\usepackage{bm}
\usepackage{color}
\usepackage{url}
\usepackage{epstopdf}
\usepackage{amsthm,mathrsfs} 
\usepackage{txfonts}
\usepackage[colorlinks=true,linkcolor=blue,urlcolor=blue,citecolor=blue,pdfusetitle]{hyperref}
\usepackage{soul}
\usepackage{lineno}

\usepackage[framemethod=TikZ]{mdframed}
    
    \usepackage{tikz}
\usetikzlibrary{calc}

\xdefinecolor{mycolor}{RGB}{62,96,111} 
\colorlet{bancolor}{mycolor}

\mdfdefinestyle{MyFrame}{%
    linecolor=black,
    outerlinewidth=2pt,
    innertopmargin=4pt,
    innerbottommargin=4pt,
    innerrightmargin=4pt,
    innerleftmargin=4pt,
        leftmargin = 4pt,
        rightmargin = 4pt,
    backgroundcolor=gray!50!white
        }

\begin{document}
\count\footins = 1000
\title{Influence of cosmological expansion in local experiments}

\author{Felix Spengler}
\affiliation{Institut f\"{u}r Theoretische Physik, Eberhard-Karls-Universit\"{a}t T\"{u}bingen, 72076 T\"{u}bingen, Germany}
\author{Alessio Belenchia}
\affiliation{Institut f\"{u}r Theoretische Physik, Eberhard-Karls-Universit\"{a}t T\"{u}bingen, 72076 T\"{u}bingen, Germany}
\affiliation{Centre for Theoretical Atomic, Molecular, and Optical Physics, School of Mathematics and Physics, Queens University, Belfast BT7 1NN, United Kingdom}
\author{Dennis R\"{a}tzel}
\affiliation{Humboldt  Universit\"{a}t  zu  Berlin,  Institut  f\"{u}r  Physik, Newtonstraße  15,  12489  Berlin,  Germany}
\author{Daniel Braun}
\affiliation{Institut f\"{u}r Theoretische Physik, Eberhard-Karls-Universit\"{a}t T\"{u}bingen, 72076 T\"{u}bingen, Germany}

\date{\today}

\begin{abstract}
Whether the cosmological expansion can influence the local dynamics, below the galaxy clusters scale, has been the subject of intense investigations in the past three decades. In this work, we consider McVittie and Kottler spacetimes, embedding a spherical object in a FLRW spacetime. We calculate the influence of the cosmological expansion on the frequency shift of a resonator and estimate its effect on the exchange of light signals between local observers. In passing, we also clarify some of the statements made in the literature. 
\end{abstract}
\maketitle

\section{Introduction}
The large scale structure of the universe is described by way of the $\Lambda$CDM cosmological model which is in accordance with current observations showing an accelerated expansion of the universe~\cite{collaboration2020planck}. The accelerated expansion is well captured by the cosmological Friedmann–Lema\^{i}tre–Robertson–Walker (FLRW) spacetime metric, once visible matter, dark matter and dark energy are accounted for in the energy density of the universe. This description is effective above the supercluster scales, where space can be considered homogeneous to a good approximation. At these scales, the evolution is dominated by the so-called Hubble flow, with structures receding from each other with relative velocities linearly proportional to their relative distance in first approximation.  

However, whether the cosmic expansion of spacetime can affect local gravitating systems has been the subject of a lively debate dating back to Einstein and Straus in the 1940s~\cite{RevModPhys.17.120,RevModPhys.18.148}. Since then, a growing body of literature has tackled the issue with sometimes conflicting predictions on the existence of local effects of the cosmological expansion~\cite{PhysRevLett.12.435,bonnor2000generalization,mashhoon2007tidal,nandra2012effect,kagramanova2006solar,carrera2006doppler,kopeikin2015optical,arakida2009time,arakida2011application,agatsuma2020expansion,giulini2014does,aghili2017effect,axenides2000some,bolen2001expansion,kerr2003standard,price2012expanding,iorio2006can,jetzer2006two,adkins2007cosmological,cooperstock1998influence,lammerzahl2008physics,RevModPhys.82.169} (see the review~\cite{RevModPhys.82.169}, and references therein, for a detailed account of the literature up to 2010). 

The major problem in addressing unambiguously which structures in the universe participate in the expansion and which do not resides in the difficulty of handling the solutions of general relativity (GR) outside extremely simplified and idealized scenarios. Ideally, one should be able to account for the local environment such as the one in the solar system and, at the same time, consider the larger environment in which the former is embedded and which, in turn, will not be described by a cosmological solution to Einstein's equations in general~\cite{bolen2001expansion}. Since such detailed description is not currently available, we are led to consider simplified scenarios with varying degrees of approximation. As highlighted in~\cite{RevModPhys.82.169}, while this is totally justified from a methodological point of view, it also demands for a conservative interpretation of the final results as indicating more likely an upper-bound on the effects of cosmic expansion on local systems than an accurate estimate. 

With these clarifications at hand, in this work we consider the effect of the global cosmological expansion on local scales using the McVittie metric~\cite{mcvittie1933mass}, describing a spherical symmetric object embedded in an expanding FLRW spacetime, and its limit case when spacetime is asymptotically de Sitter. {We will focus on the effect of the cosmic expansion on the frequency of propagating light signals.} In particular, we consider the effect of the expansion on the frequency shift of a resonator moving along different trajectories. 
We also briefly review the effect of the expansion on the exchange of light signals between different observers and clarify some statements present in the literature in this regard.

{This investigation is motivated by the rapid development of optical clocks. The great advancements in the field of optical clocks~\cite{RevModPhys.87.637,hinkley2013atomic,nemitz2016frequency,chou2010frequency,takamoto2005optical,ichiro2015cryogenic,nicholson2015systematic,mcgrew2018atomic,bothwell2019jila,PhysRevLett.123.033201} in the past 20 years -- gaining about five orders of magnitude in accuracy -- open potential new windows of exploration of fundamental physics allowing to measure time and frequency with unprecedented precision. Just this year, a measurement of the frequency ratio between three atomic clocks with a fractional frequency uncertainty below $8\times 10^{-18}$ was reported in~\cite{collaboration2021frequency}
and fractional stability of optical clocks to one part in $10^{18}$~\cite{bothwell2019jila} or even in $10^{19}$~\cite{PhysRevLett.120.103201} over average times of hundreds of seconds have been reached. Moreover, new concepts like nuclear clocks are being explored which promise even better frequency standards. {It should be noted that in SI units the current value of the Hubble parameter is around $10^{-18}$~s$^{-1}$ such that relative frequency shifts over one second -- e.g. in a space-bound cavity or Doppler measurements -- linearly proportional to it (if any) would be comparable to the current optical clocks' uncertainty, when averaging for $\sim 10^2$ seconds.}
It is then intriguing to investigate if the effect on local systems of the cosmic expansion can be of this order of magnitude, and whether current or near future experiments employing quantum technological platforms could have any hope to detect such effects.}

The work is organized as follows: In Sec.~\ref{sec:metric}, we discuss the model of spherical symmetric object embedded in an expanding FLRW spacetime that we use in the rest of the work, as well as the different observer fields we consider. Sec.~\ref{resonator} briefly reviews the derivation of the frequency shift in a resonator in curved spacetime~\cite{ratzel_frequency_2018}. Here we show how the cosmic expansion affects the resonator, depending on its trajectory. In Sec.~\ref{redshift}, we use the previous results to clarify some aspects of the imprint of the global cosmological expansion on the kinematic effects related to the exchange of light signals between different observers. {Section~\ref{acceleration} presents estimates of the magnitude of the effects previously discussed.} Finally, in Sec.~\ref{conclusions} we conclude with a discussion of our results and outlooks.  

\section{Expanding universe with a spherical inhomogeneity}\label{sec:metric}
In the existing literature, several techniques and approaches have been used to study the impact of the cosmological evolution on local systems. Beyond using perturbation theory and improved Newtonian calculations, exact solutions to the Einstein's equations have been found that describe an inhomogeneity embedded in an expanding FLRW spacetime. As discussed in detail in~\cite{RevModPhys.82.169}, two alternatives have been investigated. The first amounts to matching two known solutions of Einstein's equations, one representing the cosmological FLRW spacetime and the other the geometry induced by the isolated inhomogeneity. This has been the basis for the Einstein-Straus vacuole solution~\cite{RevModPhys.17.120,RevModPhys.18.148,schucking1954schwarzschildsche}. The second requires finding exact solutions of Einstein's equations, with the only constraint of approximating each of the two known solutions of interest in some region.  

In this work, we will follow the second strategy and consider it as a viable approach for describing a local system embedded in an expanding spacetime whose metric is the so called McVittie metric. Firstly derived in the early '30s~\cite{mcvittie1933mass}, the McVittie metric is a spherically symmetric solution to Einstein's equations and describes a non-charged, non-rotating compact object in an expanding cosmological FLRW spacetime. As such, the McVittie metric reduces, by construction, to the exterior Schwarzschild solution at small radii and to FLRW asymptotically. We restrict ourselves to the case in which the FLRW asymptotic metric describes a spatially flat spacetime, in accordance with current cosmological observations.
The analytical properties of the McVittie solution were carefully analyzed in~\cite{nolan1998point,nolan1999pointb,nolan1999pointc,nolan2014particle} where also the properties of the timelike and lightlike geodesics of the metric are considered\footnote{In~\cite{nolan1998point,nolan1999pointb,PhysRevD.81.043521,RevModPhys.82.169}, the singularity properties of the McVittie spacetime are considered. In the following, we work always far away from the Schwarschild radius ($r\gg r_S$) and thus do not concern ourselves with such issues.}.

In the following, we use mainly two coordinate representations for the McVittie metric, always assuming to be at distances from the central object much larger than its Schwarzschild radius. {We also set $c=G=1$ unless otherwise stated.} In isotropic spherical coordinates, the McVittie metric reads
\begin{equation}\label{isot}
   ds^2 =-\frac{\Bigg(1-\frac{m(t)}{2r}\Bigg)^2}{\Bigg(1+\frac{m(t)}{2r}\Bigg)^2}dt^2+\Bigg(1+\frac{m(t)}{2r}\Bigg)^4 a(t)^2(dr^2+r^2d\theta^2+r^2\sin^2\theta d\phi^2)
\end{equation}
where we are using the $(-+++)$ signature. Here, $a(t)$ indicates the scale factor of the asymptotic FLRW metric. As discussed in~\cite{RevModPhys.82.169} and references therein, the matter content of the McVittie spacetime is assumed to consist of a perfect fluid moving along the integral curves of the (normalized) vector field $\partial_t$. 
{Following~\cite{RevModPhys.82.169}, from the Einstein's equations we have $m(t)=m_0/a(t)$ with $m_0=r_S/2$ the mass of the central object\footnote{In physical units, $m_0=GM/c^2=r_S/2$, where $G$ is the gravitational constant, $c$ is the speed of light, and $M$ is the mass of the central object.} and $r_S$ its Schwarzschild radius}

A second set of coordinates that will turn out to be useful are the areal radius coordinates. The areal radius is defined as 
\begin{equation}
    R(t,r)=\left(1+\frac{m(t)}{2r}\right)^2 a(t)r.
\end{equation}
We can then adopt the change of coordinates $t\to t$, $r\to R$ and rewrite the metric, in the region $R>2m_0$, in areal radius coordinates as
\begin{equation}\label{areal}
  ds^2 =-\left(1-2\mu(R)-h(R,t)^2\right)dt^2-\frac{2h(R,t)}{\sqrt{1-2\mu(R)}}dt\,dR+\frac{1}{1-2\mu(R)}dR^2+R^2 d\theta^2+R^2\sin^2\theta d\phi^2,
\end{equation}
where $\mu(R)=m_0/R$, $h(R,t)=H(t)R$, and $H(t)=a'(t)/a(t)$ -- where the prime indicates derivative with respect to the coordinate time -- is the Hubble parameter as usual.

{Before proceeding it should be noted that, considering the current estimates for the value of the Hubble parameter at the current time $H_0\sim 70$\, s$^{-1}$\,Km\,Mpc$^{-1}$ $\sim 2\times 10^{-18}$\,s$^{-1}$ in the $\Lambda$CDM paradigm (cf. Appendix~\ref{app:lambdaCDM}), $H'_0$ is of the same order of magnitude as $H_0^2$. Thus, in the following we will consider terms in $H'$ as quadratic corrections in the Hubble parameter.}

\subsection{Limiting case: Kottler spacetime}
From the form of the metric in eq.~\eqref{isot}, it is immediate to see that, for $m_0\to 0$, we recover the FLRW metric in spherical isotropic coordinates while imposing $a(t)=1$, we obtain the exterior Schwarzschild metric. Furthermore, from the form of the metric in eq.~\eqref{areal} it is also immediate to see that, imposing the Hubble parameter to be constant $H(t)=H_0$ -- where $H_0$ is the so-called Hubble's constant -- or equivalently, choosing $a(t)=e^{H_0 t}$, we recover the line element of Schwarzschild-de Sitter (or Kottler~\cite{kottler1918physikalischen}) spacetime with cosmological constant $\Lambda=3H_0^2$ in areal radius coordinates. 

The Schwarzschild-de Sitter (SdS) case will be of relevance in the following. The SdS metric has been used in the existing literature to investigate the effect of the cosmological constant on the local dynamics in a variety of situations~\cite{islam1983cosmological,axenides2000some,kagramanova2006solar,lammerzahl2008physics} and has also been generalized to include a rotating, axis-symmetric central object, which yields the Kerr-de Sitter metric~\cite{stuchlik2004equatorial,kraniotis2004precise,PhysRevD.81.044020,kerr2003standard}. While the SdS metric encodes only the effect of the cosmological constant, it nonetheless allows for analytical solutions where only numerics can be used with the general McVittie line element. We will thus resort to the SdS line element for some of the results in the following. 

Before moving on, let us notice that SdS spacetime is static, and indeed the metric can be rewritten in the time-independent, diagonal  form\footnote{{{This form of the metric can be obtained from eq.~\eqref{areal}, with the condition $H(t)=H_0$ constant, by performing the change of coordinates $t\to t+u(R)$ with $u'(R)=H_0 R/\left(\sqrt{1-2\mu(R)}\alpha(R)\right)$ as described e.g. in~\cite{nolan2014particle}.}}}~\cite{nolan2014particle}
\begin{equation}\label{SdS}
    ds^2=-\alpha(R) dt^2+\alpha(R)^{-1}dR^2+R^2(d\theta^2+\sin^2\theta d\phi^2),
\end{equation}
where $\alpha(R)=1-r_S/R-H_0^2R^2$. We will refer to this in the following as using ``manifestly static'' coordinates. 

\subsection{Observer fields and the proper detector frame\label{subsec:observerfields_DF}}
In the next section, we are going to consider the frequency shift induced by the cosmological expansion in a resonator {attached to a support moving along}
a given trajectory in spacetime. It is thus useful to specify which timelike trajectories we are going to consider in the following. 

Independently of the specific observer field, in order to determine the frequency shift of the resonator, we will employ the metric and the Riemann tensor expressed in the proper detector frame~\cite{PhysRevD.17.1473}. The proper detector frame is defined, up to spatial rotations and with respect to a time-like trajectory $\gamma$, as the Fermi-Walker transported orthonormal tetrad  $\left\{\mathbf{e}_{\alpha}\right\},\,\,\alpha\in\{0,1,2,3\}$ with $\mathbf{e}_0=\dot{\gamma}$ the normalized four-velocity along the trajectory, that is
\begin{equation}\label{FW}
    \begin{cases}
      \mathbf{e}_0=\dot{\gamma}\\
      0=\frac{D_F \mathbf{e}_a}{\text{ds}}\equiv\frac{D \mathbf{e}_a}{\text{ds}}-\left(\mathbf{e}_a,\frac{D \mathbf{e}_0}{\text{ds}}\right)\mathbf{e}_0+(\mathbf{e}_a,\mathbf{e}_0)\frac{D \mathbf{e}_0}{\text{ds}},\,\,\forall a\in\{1,2,3\}
    \end{cases}\,
\end{equation}
where $DX/ds=\mathbf{e}_0^{\mu}\nabla_\mu X$ is the covariant derivative of the Levi-Civita connection along the direction of $\mathbf{e}_0$ and $D\mathbf{e}_0/ds=\bm{a}$ its 4-acceleration.
Due to the Fermi-Walker transport, the proper detector frame is said to be \textit{non-rotating} and can be physically realized by an observer carrying along a clock defining time and a system of three gyroscopes with spin vectors orthogonal to each other defining the spatial reference frame~\cite{misner1974gravitation}. 
Note that Fermi-Walker transport along a geodesic corresponds to parallel transport. 

We are now in the position to consider several different observer fields which will be used in the following.
\subsubsection{Cosmological observer}
The first observer field that we consider is obtained normalizing the $\partial_t$ vector field in isotropic spherical coordinates~\eqref{isot}. It is then given by 
\begin{equation}\label{velco}
    \mathbf{u}=\|\partial_t\|^{-1}\partial_t.
\end{equation}
As we commented above, the perfect fluid matter content of McVittie spacetime moves along the integral lines of such an observer field. While in FLRW such a field is geodesic, this is not the case in McVittie (or SdS) spacetime. The cosmological observer corresponds to an observer at a constant coordinate radius $r$ and, in the asymptotic region approximating FLRW, defines the so-called Hubble flow. 

The 4-acceleration $\bm{a}=\nabla_{\mathbf{u}}\mathbf{u}$ of this field is given by 
\begin{equation}
\bm{a}
=\left\{0,\frac{ \frac{m(t)}{ r^2 a^2(t)}}{\left(1-\frac{m(t)}{2 r }\right)\left(1+\frac{m(t)}{2 r }\right)^5},0,0\right\},  
\end{equation}
i.e., purely radial and outward pointing in these coordinates at the fixed value of $r$.

\subsubsection{Kodama observer}
The Kodama observer field is the normalized version of the Kodama vector field, which is a naturally distinguished field in spherically symmetric spacetimes. Indeed, the Kodama vector field ($\mathbf{v}_{\rm K}$) is the unique (up to a sign) spherically symmetric vector field orthogonal to the gradient of the areal radius. Thus, upon normalization, we obtain a naturally distinguished observer field $\mathbf{u}_{\rm K}=\|\mathbf{v}_{\rm{K}}\|^{-1}\mathbf{v}_{\rm{K}}$ corresponding to an observer at a constant areal radius. 

As discussed in detail in~\cite{RevModPhys.82.169,carrera2010geometrical}, to which we refer the interested reader for further details, the integral curves of the Kodama observer are worldlines which ‘stay’ at constant areal radius and are orthogonal to the orbits of the ${\rm SO(3)}$ isometry group. These are the key properties of the Kodama observer field, making it the natural substitute for a timelike Killing field in an arbitrary spherically symmetric spacetime. This holds true in McVittie spacetime where, in general, no timelike Killing vector field is present. Furthermore, the Kodama vector field coincides with the timelike Killing vector field in the limit in which McVittie reduces to the SdS spacetime.   

Starting from eq.~\eqref{isot}, and using the fact that $m(t)=m_0/a(t)$, the Kodama vector field is given as
\begin{equation}\label{velkod}
    \mathbf{v}_{\rm{K}}=\partial_t-H(t)R\|\partial_r\|^{-1}\partial_r.
\end{equation}
However, due to the properties of the Kodama vector field, it is convenient to work with it in areal radius coordinates eq.~\eqref{areal} in which the Kodama vector field assume the simple form $\mathbf{v}_{\rm K}=\{1,0,0,0\}$. The Kodama observer field is finally obtained as 
\begin{equation}
    \mathbf{u}_{\rm K}=\frac{1}{\sqrt{1-2\mu(R)-h(R,t)^2}}\{1,0,0,0\}.
\end{equation}
It should be noted that, the Kodama observer field is, in general, not geodesic similarly to the cosmological observer field in McVittie spacetime. The 4-acceleration can be easily computed by fixing, without loss of generality, $\theta=\pi/2$ and is given by
\begin{align}
\bm{a}_{\rm K}&=\left\{\frac{R H(t) \left(R H'(t)-\frac{(\mu(R)-h^2(R,t))(1-2\mu(R)-h^2(R,t))}{R \sqrt{1-2\mu(R) }}\right)}{(1-2\mu(R)-h^2(R,t))^2},\frac{m_0}{R^2}-\frac{R \sqrt{1-2\mu(R)} H'(t)}{1-2\mu(R)-\,h^2(R,t)}-R H(t)^2,0,0\right\},
\end{align}
for the observer at fixed areal radius $R$.

In FLRW, the Kodama observer would still be not geodesic, contrary to the cosmological one and, given the fact that the expression for the areal radius reduces to $R=a(t)r$, would correspond to an observer at a constant proper distance from the origin of the coordinate system. In the limit in which $a(t)=1$, the Kodama observer is just the stationary one of Schwarzschild spacetime.

\subsubsection{Geodesic observers}
Among the better physically justified observers to consider there are undoubtedly geodesic observers, i.e., inertial observers in free-fall. Indeed, the previous two observer fields require a proper acceleration for a spacecraft to keep on moving along their integral curves, in contrast to geodesic ones. 

Timelike geodesics in McVittie spacetime have been carefully analyzed in~\cite{nolan2014particle}. Unfortunately, for a general geodesic there are no analytical expressions and also finding the associated proper detector frame analytically is a tall order.  
Thus, in the following, when speaking of geodesic observers, we will consider the SdS limit of McVittie spacetime and work in manifestly static coordinates~\eqref{SdS}. In the case of SdS, analytical expressions for timelike geodesics have been derived~\cite{PhysRevD.78.024035} albeit involving hyper-elliptic integrals. Since we consider the SdS spacetime metric, our conclusions for what regards geodesic observers will be concerned with the local effects of a cosmological constant.

As discussed in~\cite{PhysRevD.78.024035}, the symmetries of SdS spacetime allow to define the two conserved quantities energy $E$ and angular momentum $L$ that, together with the normalization condition for the 4-velocity $\dot{\gamma}$ of a timelike ($\epsilon=1$) or null ($\epsilon=0$) geodesic trajectory, uniquely characterize the 4-velocity of the trajectory as 
\begin{equation}\label{geodSdS}
   \dot{\gamma}= \left\{\frac{E}{\alpha(R)},\sqrt{E^2-\alpha(R)\left(\epsilon+\frac{L^2}{R^2}\right)},0,\frac{L}{R^2}\right\}
\end{equation}
where we are working in static coordinates, and we have restricted ourselves, without loss of generality, to motion in the equatorial plane ($\theta=\pi/2$). 

In the following, we focus on circular and radial geodesics. The former give a crude approximation of the physical motion of planets in the solar system, while the latter describe the motion of an infalling or outwards escaping spacecraft. Radial geodesics are easily determined by fixing $L=0$ and are parametrized by the energy $E$. The corresponding Fermi transported tetrad that defines the proper detector frame is derived in Appendix~\ref{appA}. Circular orbits are instead obtained by demanding that {$\dot{R}=0$ and $\ddot{R}=0$, where the dot stands for derivative with respect to the proper time along the geodesic, and we parametrize the trajectory as $\gamma(\tau)=\left(t(\tau),R(\tau),0,\phi(\tau)\right)$ with $\tau$ the proper time. Also for this case, the Fermi transported tetrad is derived in Appendix~\ref{appA}.} {Furthermore, the results for the circular geodesics can be easily generalized to include a central spinning object, i.e., working with the Kerr-de Sitter spacetime metric. We report this case in Appendix~\ref{B5} for the interested reader.}

\section{Frequency shift for a local resonator}\label{resonator}
In~\cite{ratzel_frequency_2018} a resonator, consisting of two mirrors connected by an elastic rod which itself is fixed to a support, affected by a curved background metric is studied. Based on this, we analyze here the effects of cosmological expansion on the frequency of a resonator on different trajectories.

\begin{figure}[t!]
\centering
\includegraphics[scale=0.5]{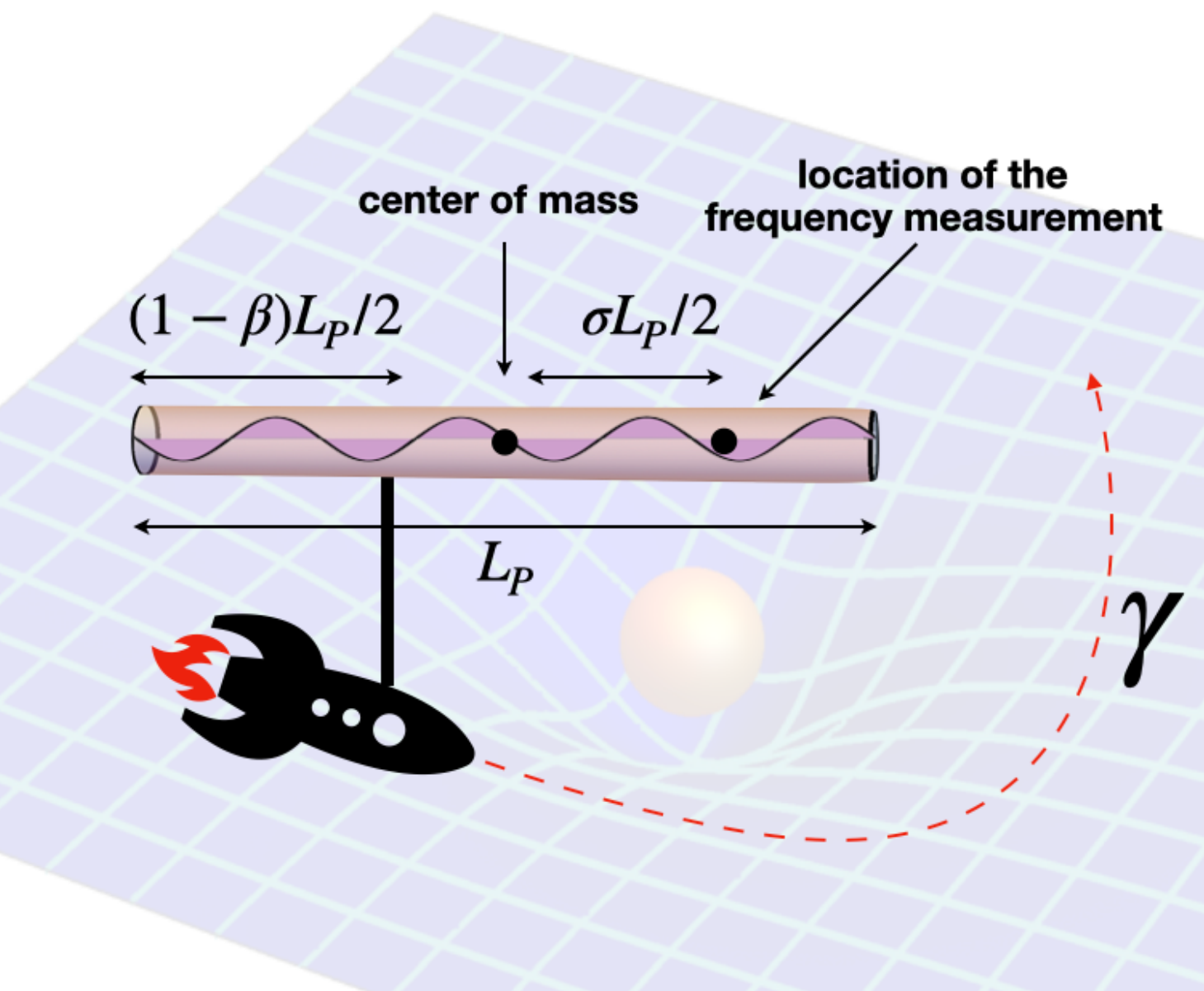}
\caption{Pictorial representation of the resonator rod carried along a trajectory $\gamma$ by an observer (the spacecraft) in a curved background. The observer trajectory $\gamma$ is the one followed by the support point at which the resonator is fixed to the observer. The proper length of the resonator is denoted by $L_p$, the support point is at a distance $\beta L_p/2$ from the center of mass and the frequency measurement is performed in an arbitrary point of the resonator at a distance $\sigma L_p/2$ from the center of mass.}
\label{res}
\end{figure}

{Attached by a support to an observer, which is characterized by a timelike trajectory and its local proper detector frame, the resonator is subject to gravitational effects of the metric (cf. Fig.~\ref{res}).}
{We indicate the spatial coordinates in the proper detector frame as $\{x,y,z\}$ along the directions defined by the spatial part of the Fermi-Walker transported tetrad $\{\bm{e}_{J}^{\mu}\}$ introduced in~\eqref{FW} and whose explicit expressions are given in Appendix~\ref{appA} for different observers\footnote{Note that, as detailed in~\cite{ratzel_frequency_2018}, the proper detector frame coordinates used for the derivation of the frequency shift are valid for distances from the point of expansion much smaller than
$
    \min \left\{c^2/|{\bf a}^J | , 1/|R^M_{NPQ}|^{1/2}, |R^M_{NPQ}|/|\partial_K R^M_{NPQ}|\right\},
$ where ${\bf a}^J=\bm{e}^{J}_{\mu}\bm{a}^{\mu}$ is the non-gravitational acceleration with respect to a local freely-falling frame while $R_{MNPQ}=\bm{e}^{\alpha}_{M}\bm{e}^{\beta}_{N}\bm{e}^{\gamma}_{P}\bm{e}^{\delta}_{Q}R_{\alpha\beta\gamma\delta}$ are the proper detector frame components of the Riemann tensor.}.} 
We assume that the resonator is aligned along the, arbitrarily chosen, $J$-direction, with $J\in \{x,y,z\}$ in the proper detector frame and that the rod's elasticity is characterized by the material's speed of sound $c_s$.

The slowly varying acceleration and tidal forces induce internal stress within the rod, which accumulates along it leading to a compression or elongation. Other contributions to the change in length are the relativistic length contraction which is subleading by a factor $c_s^2/c^2$,  the effects resulting from transverse proper acceleration, and change in trajectory of the light pulse which are second order effects in the perturbation of the proper detector frame metric. The proper acceleration of the mirrors can be ignored if the mirrors are considered lightweight compared to the rod. Additional effects of tidal acceleration in transverse directions are negligible for a slim rod. {Following the detailed derivation in~\cite{ratzel_frequency_2018},} the change in length translates to a shift in the resonance frequency which, in quadratic order of the metric perturbation and in the limit of a slowly moving observer, is given {in eq.(29) of~\cite{ratzel_frequency_2018} by}
{
\begin{equation}\label{fshift}
\frac{\Delta \omega}{\omega} \approx \frac{{\bf a}^J}{2c^2} \left( \frac{c^2}{c_s^2} \beta - \sigma\right) L_p + \frac{R_{0J0J}}{24}\left( 2 \frac{c^2}{c_s^2}(3\beta^2+1) - 3 \sigma^2 - 6\sigma\beta+1\right) L_p^2,
\end{equation}
{where $\omega$ is the resonance frequency of the oscillator in the absence
of curvature and acceleration,} ${\bf a}^J\equiv\bm{e}^{J}_{\mu}\bm{a}^{\mu}$ and $R_{0J0J}=\bm{e}^{\alpha}_{0}\bm{e}^{\beta}_{J}\bm{e}^{\gamma}_{0}\bm{e}^{\delta}_{J}R_{\alpha\beta\gamma\delta}$ are the proper acceleration and the Riemann curvature tensor components in the detector frame, $\beta\, L_p/2$ is the distance of the rod's support from the center of mass and $ \sigma\, L_p/2$ is the distance of the point of measurement from the center-of-mass (cf. Fig.~\ref{res}). Note that we have re-inserted the speed of light for clarity of exposition.} 
By design, i.e., by the choice of alignment, the acceleration and curvature parallel to the axis of the resonator are the only components to contribute in leading order to the frequency shift of the resonator. Analytic formulae for the relevant proper acceleration and curvature tensor elements and the impact of cosmological expansion on these are given for some relevant observer field in the following.

\subsection{Cosmological observer}
For the cosmological observer, the spatial component of the 4-acceleration in the proper detector frame is given by
\begin{equation}\label{eq:propacc_cosobs}
    \mathbf{a}=\left(\frac{\frac{ m(t)}{a(t)r^2}}{\left(1-\frac{m(t)}{2r}\right) \left(1+\frac{m(t)}{2r}\right)^3},0,0\right).
\end{equation}
An expansion to second order in $ r a'(t)/a(t) $ and first order in $ r^2 a''(t)/a(t)$ (here, derivatives are taken with respect to coordinate time) of the relevant terms of the Riemann curvature tensor in the proper detector frame results in
\begin{align}
   R_{0x0x} &\approx  -\frac{2 m(t)}{r^3 a(t)^2 }\frac{1}{\left(1+\frac{m(t)}{2 r}\right)^6} + \frac{m(t) }{r}\frac{1}{1-\frac{m(t)}{2 r}} \left(\frac{a'(t)}{a(t)}\right)^2 - \frac{\left(1+\frac{m(t)}{2 r}\right)}{\left(1-\frac{m(t)}{2 r}\right)}\frac{a''(t)}{a(t)},
   \\
   R_{0y0y} = R_{0z0z} & \approx  \frac{ m(t)}{r^3 a(t)^2 }\frac{1}{\left(1+\frac{m(t)}{2 r}\right)^6} + \frac{m(t) }{r}\frac{1}{1-\frac{m(t)}{2 r}} \left(\frac{a'(t)}{a(t)}\right)^2 - \frac{\left(1+\frac{m(t)}{2 r}\right)}{\left(1-\frac{m(t)}{2 r}\right)}\frac{a''(t)}{a(t)},
 \end{align}
 where we see that in the radially aligned $\bm{e}^\mu_x$ direction the curvature is different from the two orthogonal directions.
Apart from the $ \left(a'(t)/a(t)\right)^2 $ and $a''(t)/a(t)$ factors, the cosmological expansion only enters as $ r a(t)$ in both the Riemann tensor and the acceleration in the Fermi-Walker transported detector frame. 
{Expanding the scale factor $a(t)\approx a(t_0)(1+H_0{\Delta t})$ gives {terms linear in $H_0$} in both the acceleration and the curvature, all suppressed by factors $m_0/r$. As we show in the following, this is not the case for the other observers that we consider, which do not follow the Hubble flow.}
{Note that, in the limit {$m(t)/r\to 0$}, i.e., when the usual FLRW metric is recovered, we remain with corrections  
{quadratic in $H_0$} in agreement with previous results in the literature~\cite{kopeikin2015optical}.}

\subsection{Kodama observer}
Similarly, starting from the McVittie metric in areal radius coordinates, we can calculate the quantities relevant to the frequency shift of a resonator on the trajectory of a Kodama observer.
The spatial component of 4-acceleration in the proper detector frame is given by
\begin{equation}\label{eq:propacc_kodama}
\mathbf{a}_{\rm K} = \left(\frac{ \frac{m_0}{R^2} - R H^2(t)}{\sqrt{1-R^2 H(t)^2-\frac{2 m_0}{R}}}  - \frac{\sqrt{1-\frac{2 m_0}{R}} R H'(t)}{\left(1-R^2 H(t)^2-\frac{2 m_0}{R}\right)^{3/2}},0,0\right).
\end{equation}
The relevant components of the Riemann tensor in the detector frame, i.e., the ones entering~\eqref{fshift}, are given by
\begin{align}\label{eq:kodama_curvature}
R_{0x0x}=&-\frac{H'(t)}{\sqrt{1-\frac{2 m_0}{R}}}-H(t)^2-\frac{2 m_0}{R^3},
\\
R_{0y0y}=R_{0z0z}=& \frac{m_0}{R^3} - H^2(t) -\frac{\sqrt{1-\frac{2 m_0}{R}} H'(t)}{1-R^2 H(t)^2-\frac{2 m_0}{R}}.
\end{align}
It should be noted that, both the proper acceleration and any component of the Riemann curvature tensor in the proper detector frame do not contain any linear term in the Hubble parameter $H$ {nor any instance of $a(t)$ from which a linear term in $H_0$ could arise when expanded for small time differences}. The same is true for the Ricci and Einstein tensors and the scalar curvature {and it is in stark contrast to the case of the cosmological observer.}
Finally, it is worth mentioning that, performing the same calculations starting from the McVittie metric in isotropic coordinates calls for some care. Indeed, one needs to impose the constancy of the areal radius, characteristic of the Kodama observer, in order to correctly account for correction to the resonator frequency and obtain results that agree with the one discussed here.

\subsection{Geodesic observers}\label{subsec:resonator_geodesic}
As discussed in the previous section, finding the proper detector frame for a general geodesic observer in McVittie spacetime is a tall order. We thus focus on geodesics in SdS spacetime working in static coordinates. 

Let us first consider an observer freely-falling along an equatorial circular geodesic. The Riemann curvature tensor in the proper detector frame is given by
\begin{align}
R_{0x0x}= & \frac{-3 \frac{r_S}{R^3} \alpha(R) \cos \left(\phi  \sqrt{4-\frac{6 r_S}{R}}\right)-H_0^2  (4 -9 \frac{r_S}{R})- \frac{r_S}{R^3}}{2  \left(2 -3 \frac{r_S}{R}\right)},
\\
R_{0y0y} = & \frac{1}{2- 3 \frac{r_S}{R}} \left( \frac{r_S}{R^3} - 2 H_0^2\right)
\\
R_{0z0z} = & \frac{3 \frac{r_S}{R^3} \alpha(R) \cos \left(\phi  \sqrt{4-\frac{6 r_S}{R}}\right)-H_0^2  (4 -9 \frac{r_S}{R})- \frac{r_S}{R^3}}{2  \left(2 -3 \frac{r_S}{R}\right)}.
\end{align}
Note that, the trigonometric functions appearing in these expressions originate from requiring the detector frame to be non-rotating (see also Appendix~\ref{appA}).

In the case of a radial, equatorial geodesic in SdS, characterized by a vanishing angular momentum $L=0$, we find for the components of interest of the proper detector frame Riemann tensor
\begin{align}
R_{0x0x} = & - \frac{r_S}{R^3} - H_0^2,
\\
R_{0y0y} = R_{0z0z} = & \frac{r_S}{2 R^3} - H_0^2.
\end{align}
Note that, these expressions coincide, at a fixed value of the areal radius, with the ones of the Kodama observer~\eqref{eq:kodama_curvature} in the limit $H(t) \to H_0$.

The previous expressions show that, for both the circular and the radial geodesics, the leading order correction to the frequency shift of the resonator resulting from the cosmological expansion is  
{ proportional to} $H_0^2$. {Let us also notice that the same conclusion can be reached for the particular case of circular geodesics in Kerr-de Sitter spacetime~\cite{carter1968hamilton,demianski1973some,gibbons2005general}, in which a central rotating body is considered (cf. Appendix~\ref{appA} for  
additional details).}

\section{Redshift and satellite tracking}\label{redshift}
Having considered the impact of the global cosmological expansion on a local experiment, we conclude with a brief overview of the effect of the same expansion on the frequency redshift of signals exchanged between observers and the related concept of double Doppler tracking (DDT). These kinematic effects have been treated in detail in the existing literature~\cite{kagramanova2006solar,lammerzahl2008physics,carrera2006doppler,RevModPhys.82.169}. Here, we focus on clarifying some of the results in the literature by following the derivation in~\cite{RevModPhys.82.169}. 

In the case of FLRW spacetime, the redshift formula for exchanges of light signals between two observers following the Hubble flow is easily obtained. Consider two cosmological observers at $r_0$ and $r_1$, respectively, in isotopic spherical coordinates\footnote{In these coordinates the FLRW line element has the usual form $ds^2=-dt^2+a^2(t)(dr^2+r^2\sin\theta^2d\phi^2+d\theta^2)$.}, then the ratio between the frequency emitted by the first observer and the one received by the second observer is given by {$\omega_1/\omega_0=a(t_0)/a(t_1)\sim 1-H_0(t_1-t_0),$}
where we have assumed the leading order of the Hubble parameter to be $H_0 ={\rm const.}$,  and where the frequency measured by an observer $\mathbf{u}$ is given by the scalar product between the observer field and the null tangent to the light signal $\mathbf{k}$, i.e. $|g(\mathbf{u},\mathbf{k})|$. 
We notice that the redshift encodes a correction linear in the Hubble constant. As shown in~\cite{RevModPhys.82.169}, this persists also in the case in which a spherical inhomogeneity is included. 

The DDT, as the name suggests, is a technique used to track the position of spacecrafts. In order to address the DDT, we need to consider the ratio between the frequency emitted by an observer and the frequency received back by the same observer after the light signal has been reflected by an arbitrarily moving ``spacecraft''. In FLRW, considering the cosmological observer field and a spacecraft reflecting back the light signal upon reception, this ratio can be broken into three terms $\omega_2(t_2)/\omega_0(t_0)=(\omega_2/\omega'_1)(\omega'_1/\omega_1)(\omega_1/\omega_0)$. Here, the observer receiving (at $t=t_2$) and sending (at $t=t_0$) the signal is at a fixed value of the coordinate radius $r$. Also notice that, for the cosmological observer {in FLRW}, proper time coincides with the coordinate time $t$. The ratio $\omega_1/\omega'_1$ represents the ratio between the frequency at the reflection point \textit{as measured by the cosmological observer at that point} and the frequency after reflection as measured by the same observer. This ratio accounts for the relativistic Doppler shift due to the motion of the spacecraft relative to the cosmological observer field at the reflection point. The other two ratios are easily obtained from the previous expression of the single-way redshift. Altogether, one arrives at eq.~(140) of~\cite{RevModPhys.82.169}
\begin{equation}\label{2wayratioFRW}
    \frac{\omega_2(t_2)}{\omega_0(t_0)}=\frac{a_0}{a_2}\left(2\frac{1-\beta_{\mathbf{u}}^{\hat{\bm{k}}}(\mathbf{v})}{1-\beta_{\mathbf{u}}(\mathbf{v})^2}-1\right).
\end{equation}
Here, we have considered a spacecraft with four velocity $\mathbf{v}$ whose relative velocity with respect to the observer field $\mathbf{u}$ at the reflection point is $\bm{\beta}_{\mathbf{u}}(\mathbf{v})=(\mathbf{v}-|g(\mathbf{v},\mathbf{u})|\mathbf{u})/|g(\mathbf{v},\mathbf{u})|$~\cite{RevModPhys.82.169}. Furthermore, we have considered a null signal propagating radially between the emitter and the reflection point with tangent $\mathbf{k}$ whose normalized projection in the rest frame of the observer is $\hat{\bm{k}}$. Finally, with $\beta^{\hat{\bm{k}}}_{\mathbf{u}}$ we indicate the projection of the relative velocity along the unit vector $\hat{\bm{k}}$ in the rest frame of $\mathbf{u}$. Eq.~\eqref{2wayratioFRW} relates the frequency shift to the spacecraft spatial velocity and can be approximated to linear order in $\beta$ and $H_0\Delta t_{20}$ with $\Delta t_{20}=t_2-t_0$ giving $\omega_2(t_2)/\omega_0(t_0)\approx 1-2\beta^{\hat{\bm{k}}}_{\mathbf{u}}(\mathbf{v})-H_0\Delta t_{20}$ showing once more a linear correction in $H_0$. It is clear from our previous discussion that the linear term in $H_0$ originates from the analogous term in the one-way redshift. 

The last step in accounting for the DDT is to  
differentiate the previous expression with respect to the proper time of the observer at reception of the reflected signal. In~\cite{RevModPhys.82.169}, this calculation is detailed, and its result is
\begin{align}\label{flrwDoubleDoppler}
  \frac{1}{\omega_0(t_0)}\frac{d\omega_2(t_2)}{dt_2}&\approx-2\left\{\alpha_{\mathbf{u}}^{\hat{\bm{k}}}(1-3\beta_{\mathbf{u}}^{\hat{\bm{k}}}-3H_0\Delta t_{20}/2)+H_0\beta_{\mathbf{u}}^{\hat{\bm{k}}}\right\},
\end{align}
where now $\alpha_{\mathbf{u}}^{\hat{\bm{k}}}$ is the relative spatial acceleration of the spacecraft trajectory with respect to the observer field in the direction of the unit vector $\hat{\bm{k}}$. It should be emphasized here that, in the rate of the DDT ratio above, not all the corrections linear in $H_0$ can be traced back to the one-way redshift. Indeed, the derivative with respect to $t_2$ is also responsible for the introduction of such corrections as can be easily seen from eq.~(143) in~\cite{RevModPhys.82.169}.

We now want to show that such linear corrections in $H_0$ are a peculiarity of the cosmological observer already in FLRW. This result then extends trivially to the case of McVittie spacetime.
A first hint of this fact is given by considering the simple case of de Sitter spacetime\footnote{See also appendix~\ref{app:red} for further details.}, and the Kodama observer field which, in the region of interest, is a timelike Killing vector field. The one way redshift is thus given by~\cite{wald2010general}
\begin{align}\label{sdsredshift}
    \frac{\omega_1}{\omega_0}&=\frac{\sqrt{\alpha(R_0)}}{\sqrt{\alpha(R_1)}}= 1+(R_1^2-R_0^2)\frac{H_0^2}{2}+\mathcal{O}\left(H_0^4 R^4\right),
\end{align}
with {$R$ the largest scale among $R_0$ and $R_1$ entering the problem and} $\alpha(R)=1-H_0^2 R^2$. Note that the same expression holds in the case of SdS spacetime, where only the functional form of $\alpha(R)$ changes (see also Sec.~\ref{acceleration} for an alternative way to compute the one-way redshift for the Kodama observer). Moreover, we point out that the expression for the redshift as expressed in~\cite{kagramanova2006solar} contains an error -- in that $R_0$ and $R_1$ are interchanged. 

The previous expression shows that no {contribution linear in  $H_0$} appears while the leading corrections are {proportional to $H_0^2$}. It also tells us immediately that the only possible source of {corrections linear in $H_0$} in the DDT rate could be the time derivative. However, no contributions of this nature arise from the time derivative. A simple way to see this is to work in static coordinates for de Sitter spacetime. In these coordinates the Kodama observer field is $\mathbf{u}_{\rm K}=\alpha^{-1/2}\partial_t$ and the proper time for such an observer coincides with the coordinate time  
{to first order in $H_0R$}. We can thus ignore the difference between the proper and the coordinate time. The DDT ratio can be obtained in complete analogy to the previous case of the cosmological observer as
\begin{equation}\label{ddt-Kod}
    \frac{\omega_2}{\omega_0}=\left\{2\frac{1-\beta_{\mathbf{u}_{\rm K}}^{\hat{\bm{k}}}(\mathbf{v})}{1-\beta_{\mathbf{u}_{\rm K}}(\mathbf{v})^2}-1\right\},
\end{equation}
where we have used the fact that $\|\mathbf{u}_{\rm K}\|_{t=t_0}/\|\mathbf{u}_{\rm K}\|_{t=t_2} =1$. Next, by using the null condition for an inward directed (assuming $R_1>R_2=R_0$) radial lightlike geodesic  
\begin{equation}
    \int_{t_1(t_2)}^{t_2}dt=-\int_{R_1(t_1(t_2))}^{R_2}\frac{dR}{\alpha},
\end{equation}
we obtain
\begin{align}
    \frac{dt_1}{dt_2}=\left(1+\beta^{\hat{\bm{k}}}_{\mathbf{u}_{\rm K}}(\mathbf{v})\right)^{-1},
\end{align}
where it can be shown that the relative spatial velocity of the spacecraft at the reflection point has the form $\beta_{\mathbf{u}_{\rm K}}^{\hat{\bm{k}}}(\mathbf{v})=\alpha(R_1)^{-1}dR_1/dt_1$ (cf. Appendix~\ref{app:red}). We then see that this derivative does not contain any correction  
{proportional to $H_0$} so that the rate of the DDT will have leading corrections quadratic in the Hubble parameter. 
{This same argument extends straightforwardly to the case of SdS spacetime (cf. Appendix~\ref{app:red}). Moreover, since SdS spacetime is a special case of the McVittie one, with $H'=0$ (cf. eq.~\eqref{eq:diag_areal}), the argument should apply for the general case showing that the DDT ratio and rate contain corrections at most quadratic in the Hubble parameter, which strongly limits the possibility to observe such effects.}

In order to strengthen our point, let us consider also a generic, freely falling observer field in SdS spacetime. In the equatorial plane, this observer field is parametrized as in eq.~\eqref{geodSdS} with $\epsilon=1$. From the form of the metric in eq.~\eqref{SdS}, and the parametrization of timelike and null geodesics in eq.~\eqref{geodSdS}, it is easy to deduce that the leading order corrections of the redshift ratio have to be at least of second order in the Hubble constant so that also a generic geodesic observer does not have access to corrections linear in {$H_0$} in the frequency ratio of the double Doppler tracking.


\section{Differential acceleration \& Expansion's effects estimate}\label{acceleration}

An intuitive understanding of the differences between the observers considered can be gained through their proper acceleration in a weak field limit and the comparison to the Newtonian limit. In eq.~\eqref{eq:propacc_cosobs} and eq.~\eqref{eq:propacc_kodama} the proper acceleration is given for the cosmological observer and the Kodama observer, respectively, while it vanishes for the geodesic observer by definition. We immediately see that eq.~\eqref{eq:propacc_cosobs} contains only an acceleration due to the central object that is, to lowest order in {$m_0/a(t)r$}, the Newtonian gravitational acceleration due to a mass $m_0$ at a distance $a(t)r$. We can conclude that the cosmological observer freely follows the Hubble flow but accelerates against the gravitational pull of the central object, which agrees with its standard interpretation. In eq.~\eqref{eq:propacc_kodama}, we find a term of lowest order in {$m_0/R$ and $HR$} that coincides with the Newtonian gravitational acceleration due to a central mass $m_0$ at distance $R$. The additional terms represent an inward acceleration that depends on the cosmological expansion. These terms result from the property of the Kodama observer to be located at constant areal radius, which implies that it is accelerating against the gravitational effect of the cosmological expansion such that it will never join the Hubble flow. 

To lowest non-trivial order, the radial proper acceleration of the Kodama observer in the corresponding proper detector frame becomes 
\begin{equation}\label{eq:propacc_kodama_rad0}
\mathbf{a}_{\rm K,R} = \frac{m_0}{R^2} - R (H^2(t) + H'(t))\,.
\end{equation}
If the Kodama observer is realized and test matter is released by it, $\mathbf{a}_{\rm K,R}$ is their differential acceleration. We recognized that the small quantities $H^2(t),H'(t)\sim H_0^2$ are multiplied by the potentially large quantity $R$. This seems like a potential opportunity for a measurement of Hubble parameter $H_0$ and the cosmological constant $\Lambda$. However, in the described setup, the fundamental challenge would be to realize the Kodama observer without knowledge of the cosmological acceleration. The only obvious possibility seems to be a measurement of the distance to the central object, which seems very challenging, in particular, for very large $R$.  

The acceleration in eq.~\eqref{eq:propacc_kodama_rad0} is equivalent, up to a sign, to the gravitational acceleration appearing in the Newtonian limit of a gravitating spherically symmetric central object in an expanding spacetime (e.g. see eq.~(87) in \cite{Nandra2012} and eq.~(1) of \cite{nandra2012effect}). The sign change is the result of $\mathbf{a}_{\rm K,R}$ being the non-gravitational acceleration necessary to compensate for the gravitational one. There exists an extended literature about the gravitational acceleration being proportional to $H^2$ due to the cosmological expansion and its local measurability~\cite{nandra2012effect,cooperstock1998influence,adkins2007cosmological,price2012expanding,axenides2000some,giulini2014does,agatsuma2020expansion,Nandra2012,RevModPhys.82.169} that we do not review here.

By expanding all expressions to second order in $HR$, we can approximately diagonalize the McVittie metric in \eqref{areal} by a redefinition of the time variable to obtain the line element (see Appendix \ref{app:perturbative} for details)
\begin{equation}\label{eq:diag_areal}
  ds^2 \approx -\left(1-\frac{2m_0}{R} - (H(t)^2 + H'(t)) R^2\right)dt^2+\left( 1 + \frac{2m_0}{R} + H(t)^2R^2\right) dR^2+R^2 d\theta^2+R^2\sin^2\theta d\phi^2\,
\end{equation}
in the weak field limit where $m_0/R \ll 1$, where we have also neglected terms proportional to $m_0 H(t)^2 R$ and $m_0 H'(t) R$. Identifying the zero-component of the perturbation of the metric, with respect to the flat Minkowski metric, with a Newtonian potential, i.e. $g_{00} = -1 - 2\Phi = -1 + 2m_0G/(c^2 R) + R^2(H^2 + H')/c^2$, leads to a gravitational redshift/time dilation proportional to $m_0(1/R_1 - 1/R_2) + (R_1^2-R_2^2)(H^2 + H')/2$ for two observers located at $R_1$ and $R_0$ \footnote{ Here, we assume that the derivatives of the components of the metric with respect to $t$ are much smaller than their derivatives with respect to $R$ to recover the stationary weak field situation (see e.g.\cite{carroll1997lecture}).}. Alternatively, this expression for the redshift can be directly deduced from the effective gravitational acceleration $-\mathbf{a}_{\rm K,R}$ by interpreting it in terms of an effective metric in the Newtonian limit.

The current value of the Hubble constant is $H_0\sim 2.2\times 10^{-18}\,\mathrm{s}^{-1}$ and that of the cosmological constant is $\Lambda=3H_0^2 \Omega_\Lambda/c^2 \sim 10^{-52}$~m$^{-2}$. Considering the $\Lambda$CDM model and neglecting the small contribution of radiation at present day, we obtain $H^2+H'= (\Lambda c^2 - H_0^2)/2$, where we have taken into account that $\Omega_\Lambda+\Omega_m = 1$ in the $\Lambda$CDM model (see appendix \ref{app:lambdaCDM} for further details). 
{Following~\cite{kagramanova2006solar}, and using eq.~\eqref{eq:diag_areal}, by assuming Earth as the central object, a satellite at an altitude of $R_1 = 15000\,\mathrm{km}\gg R_0$, and assuming a clock comparison accuracy of $10^{-15}$ and no deviation in the frequency redshift with respect to the prediction of Einstein's theory without cosmological expansion, we can estimate an upper bound to the cosmological constant} of $|\Lambda| \lesssim 2\times 10^{-29}$\,m$^{-2}$ (compare with~\cite{kagramanova2006solar}) assuming precise knowledge of $H_0^2$, $m_0$ and $R_2$ from other measurements. {This translates to an upper bound on $H_0^2$, based on the same parameters, of $H_0^2 \lesssim 10^{-12}$\,s$^{-2}$ assuming $\Lambda$, $m_0$ and $R_2$ to be known precisely}. If we consider instead the Sun as a central object, a relative clock accuracy of $10^{-19}$, and a spacecraft at a distance from the Sun comparable to the one of the Voyager~1, i.e. $R_1\sim 23\times 10^{12}$~m, we can push the bounds to $|\Lambda| \lesssim 10^{-45}$\,m$^{-2}$ and $H_0^2 \lesssim 3\times 10^{-29}$\,s$^{-2}$, ``only'' seven orders of magnitude away from the currently accepted values. Considering a scaling of the clock uncertainty with the inverse of the square root of the averaging time~\cite{RevModPhys.87.637} and the fact that currently an averaging time of the order of $10^2$\,s is needed to reach an uncertainty of the order of $10^{-19}$, we see that to fill the six orders of magnitude gap would require around $10^6$ years of integration time. It is thus clear that further advances in clock accuracy are needed to be able to assess cosmological quantities in this kind of local experiments.

In addition to redshift measurements, another option to estimate the non-Newtonian gravitational acceleration due to the cosmological expansion would be, for example, its accumulated effect on a spacecraft which could be measured, in principle, by Doppler tracking the time evolution of the spacecraft's velocity. However, the basic mechanism would be a frequency comparison which should result in similar fundamental limits as we have found for gravitational redshift measurements above.
The situation becomes even worse if one tries to measure the Hubble or cosmological constant through the frequency shift of a resonator, as described in sec. \ref{resonator}. For a geodesic observer in SdS (see sec. \ref{subsec:resonator_geodesic}), assuming the value of $H_0\sim 2.2\times 10^{-18}\,\mathrm{s}^{-1}$, and a speed of sound $c_s = 5000$\,m/s (comparable to that of aluminium) the relative frequency shift of a resonator of length $10$\,m is only $\sim 10^{-42}$, which is many orders of magnitude away from measurability.


\section{Discussion}\label{conclusions}
In the first part of this work, we have investigated the impact of the global cosmological evolution on a very local experiment, i.e., on the frequency shift in an optical resonator. In order to model something akin to a local inhomogeneous environment immersed in an expanding universe, we have considered the McVittie metric. This metric describes a spherical object embedded in a FLRW {cosmological spacetime and, together with its limiting case of Schwarzschild-de Sitter spacetime, has been largely used in the literature exploring local effects of the cosmological expansion (cf.~\cite{RevModPhys.82.169} and references therein). A word of caution is in order here to correctly interpret the results obtained in this work, as well as in the existing literature. As already discussed in the introduction -- and highlighted in, e.g.,~\cite{bolen2001expansion,RevModPhys.82.169} -- to fully address the problem of the local effects of global expansion in General Relativity it would be necessary to model the hierarchy of embedded systems, from the Solar system to the cosmological solution passing through Galaxies and Cluster scales, at the level of at least controlled approximations to exact GR's solutions. While this is currently a tall order, resorting to (crude) approximations like McVittie spacetime and the SdS spacetime can guide us in obtaining estimates of the effects of global expansion in local systems. The caveat is that such estimates, like the ones obtained in this work, have to be interpreted as upper bounds to the effects of interest since, in realistic systems, the effects of the cosmological expansion would be further obscured by the growing complexity of local structures.}

{With these considerations at hand, we}{ have considered the shift in the frequency of {an optical resonator}, moving on various trajectories in McVittie and SdS spacetimes, due to the cosmic expansion. We have shown that  
{this frequency shift is proportional to} $H_0^2$ {in the case of freely-falling observers as well as the Kodama observer field. The former are physically interesting since they do not require the knowledge of the underlying spacetime to be realized. The latter, i.e. the Kodama observer field, an observer at a constant value of the areal radius, is instead geometrically singled-out in spherically symmetric spacetimes.} {Linear terms appear when considering observers following the Hubble flow -- which however makes them unpractical for local experiments.}
This is in accordance with the results in the existing literature, where several other effects -- from light bending to perihelion precession -- have been investigated leading to corrections with the same  
{proportionality to the square of} the Hubble parameter. Despite the smallness of the current value of $H_0$ that casts these effects outside current technological possibilities, similarly to all other effects studied previously, they nonetheless show the imprint that the cosmic expansion can have on localized systems and their dynamics. }

In the last part of our work, we have reviewed the effect of the cosmic expansion on the redshift ratio and double Doppler tracking in order to clarify some claims present in the literature. In particular, we have shown that linear corrections $H$ to the redshift and DDT emerge when considering the cosmological observer but are not otherwise present in general. We have argued that the general result consists of corrections that are at least quadratic in $H$, placing these effects on the same footing as the other effects discussed in the literature.

In conclusion, despite the high degree of idealization of the local gravitational environment entailed by the McVittie or SdS metric, the expansion of the universe is able to affect local experiments albeit in a way that places its detection beyond current technological capabilities.

\section*{Acknowledgements}
The authors wish to thank Domenico Giulini for inspiring conversations.
A.  Belenchia and D. Braun acknowledge  support  from the Deutsche Forschungsgemeinschaft (DFG, German Research Foundation) project number BR 5221/4-1. D. R\"{a}tzel thanks the Humboldt Foundation and the Marie Sklodowska-Curie Action IF program (``Phononic Quantum Sensors for Gravity'' grant number 832250 — PhoQuS-G) for support.

\bibliography{references2.bib}

\appendix

\section{$\Lambda$CDM}\label{app:lambdaCDM}
In the $\Lambda$CDM model, the scale factor of the FLRW spacetime is given by
\begin{equation}
    a(t) = \left(\frac{1-\Omega_\Lambda}{\Omega_\Lambda}\right)^{1/3} \sinh^{2/3}\left(\frac{t}{t_\Lambda}\right),
\end{equation}
where $t_\Lambda = 2/\left(3 H_0 \sqrt{\Omega_\Lambda}\right)$ is the cosmological timescale, the contribution $\Omega_{\rm rad}$ to the total matter content of the universe was considered to be negligible, and we used the approximation $\Omega_m+\Omega_\Lambda=1$.
For the Hubble parameter we obtain
\begin{equation}
    H(t) = \frac{\dot{a}(t)}{a(t)} = H_0 \sqrt{\Omega_\Lambda} \coth \left(\frac{t}{t_\Lambda}\right),
\end{equation}
where $H_0=H(t_0)$ is the current value of the Hubble parameter at the present age of the universe $t_0,$ such that $ a(t_0) = 1$.
The leading order of the non-Newtonian contribution to the curvature of McVittie spacetime  from eq.~\eqref{eq:kodama_curvature} is then given by
\begin{equation}
    -(H^2(t_0)+H'(t_0)) = \frac{1}{2}(H_0^2-\Lambda).
\end{equation}

\section{Proper detector frames: Fermi-Walker transported tetrads for various observers}\label{appA}
As discussed in the main text, the proper detector frame can be defined along any timelike worldline as an orthonormal tetrad whose timelike element coincides with the normalized tangent to the worldline and the remaining orthogonal spacelike elements are Fermi-Walker transported along the worldline. In other terms, given a worldline $\gamma$ whose normalized tangent we call $\dot{\gamma}$, the proper detector frame is defined via 
\begin{equation}
    \begin{cases}
      \mathbf{e}_0=\dot{\gamma}\\
      0=\frac{D_F \mathbf{e}_a}{\text{ds}}\equiv\frac{D \mathbf{e}_a}{\text{ds}}-\left(\mathbf{e}_a,\frac{D \mathbf{e}_0}{\text{ds}}\right)\mathbf{e}_0+(\mathbf{e}_a,\mathbf{e}_0)\frac{D \mathbf{e}_0}{\text{ds}},\,\,\forall a\in\{1,2,3\}.
    \end{cases}\,
\end{equation}
solving the system of coupled differential equations defined above for a general timelike curve in McVittie spacetime is a tall order. Other methods for determining the proper detector frame have also been developed~\cite{maluf2008construction,klein2008general} which however do not alleviate the problem of finding analytical expressions for McVittie spacetime. Nonetheless, the cases considered in the main text are such that the Fermi-Walker (FW) transported tetrad can actually be constructed by solving the above system of differential equations. This holds for the cosmological and Kodama observer in a general McVittie spacetime, and for radial and circular geodesic observers in SdS spacetime. 

In this appendix we report the explicit form of the tetrads for completeness.

\subsection{Cosmological observer in McVittie spacetime}
The FW transported tetrad for the cosmological observer in McVittie spacetime in isotopic coordinates and in matrix form where each line is one of the tetrad vectors $\mathbf{e}_J$ reads
\begin{equation}
   \mathbf{e}_J^\mu=\left(
\begin{array}{cccc}
 \frac{\left(1+\frac{m(t)}{2 r}\right)}{\left(1-\frac{m(t)}{2 r}\right)} & 0 & 0 & 0 \\
 0 & \frac{1}{a(t) \left(1+\frac{m(t)}{2 r}\right)^2} & 0 & 0 \\
 0 & 0 & \frac{1}{r a(t) \left(1+\frac{m(t)}{2 r}\right)^2} & 0 \\
 0 & 0 & 0 & \frac{1}{r a(t) \left(1+\frac{m(t)}{2 r}\right)^2} \\
\end{array}
\right).
\end{equation}
This tetrad is trivially obtained. Indeed, the spatial unit vectors $\mathbf{e}_r,\,\mathbf{e}_\theta,\, \mathbf{e}_\phi$ are just the normalized versions of the vectors $\partial_r,\,\partial_\theta$ and $\partial_\phi$ respectively. 

\subsection{Kodama observer in McVittie spacetime}
The FW transported tetrad for the Kodama observer in McVittie spacetime in areal radius coordinates and in matrix form where each line is one of the tetrad vectors reads

\begin{equation}
    \mathbf{e}_J^\mu=\left(
\begin{array}{cccc}
 \sqrt{\frac{1}{1-2\mu(R)-h^2(R,t)}} & 0 & 0 & 0 \\
 {-} \frac{h(R,t)}{\sqrt{(1-2 \mu(R)) \left(1-2\mu(R)-h^2(R,t)\right)}} & {\sqrt{1-2\mu(R)-h^2(R,t)}} & 0 & 0 \\
 0 & 0 & \frac{1}{R} & 0 \\
 0 & 0 & 0 & \frac{1}{R\sin\theta} \\
\end{array}
\right).
\end{equation}
This tetrad is easily obtained.: The spatial unit vectors $\mathbf{e}_\theta,\, \mathbf{e}_\phi$ are just the normalization of the vectors $(0,0,1,0)$ and $(0,0,0,1)$, respectively. For what concerns $\mathbf{e}_R$, it is easily found by just imposing $g(\mathbf{e}_0,{\mathbf{e}_R})=0$ and $g(\mathbf{e}_R,{\mathbf{e}_R})=1$ with an ansatz $\mathbf{e}_R=(v_0,v_R,0,0)$.

\subsection{Radial geodesic observer in SdS spacetime}
The FW transported tetrad for the radial geodesic observer in SdS spacetime in manifestly static coordinates and in matrix form where each line is one of the tetrad vectors is given by

\begin{equation}
   \mathbf{e}_J^\mu= \left(
\begin{array}{cccc}
 \frac{E}{\alpha(R)} & \sqrt{E^2-\alpha(R)} & 0 & 0 \\
 \frac{\sqrt{E^2-\alpha(R)}}{\alpha(R)} & {E} & 0 & 0 \\
 0 & 0 & \frac{1}{R} & 0 \\
 0 & 0 & 0 & \frac{1}{R \sin\theta} \\
\end{array}
\right).
\end{equation}
This tetrad is easily obtained: The spatial unit vectors $\mathbf{e}_\theta,\, \mathbf{e}_\phi$ are just the normalization of the vectors $(0,0,1,0)$ and $(0,0,0,1)$ respectively. For what concerns $\mathbf{e}_R$, it is easily found by just imposing $g(\mathbf{e}_0,{\mathbf{e}_R})=0$ and $g(\mathbf{e}_R,{\mathbf{e}_R})=1$ with an ansatz $\mathbf{e}_R=(v_0,v_R,0,0)$.

\subsection{Circular orbit geodesic observer in SdS spacetime}
In the case of a circular geodesic orbits, deriving the FW transported tetrad turns out to be more demanding than in the previous cases. Following~\cite{brito2020synchrotron}, imposing the condition $\dot{R}=0$ in eq.~\eqref{geodSdS} allows to fix the value of the conserved energy $E$ and subsequently imposing the 4-acceleration to be vanishing fixes the conserved angular momentum as 
\begin{align}
    &E^2=R\frac{\alpha(R)^2}{R-3r_S/2}\\
    &L^2=R^2\frac{r_S-2H_0^2R^3}{2R-3r_S}.
\end{align} 
It should be noted that in SdS spacetime circular geodesics exist in the region $3r_S/2<R<(r_S/2H_0)^{1/3}$. 

At this point, we should notice that the vector $\tilde{\mathbf{e}}_{\theta}=\|g_{\theta\theta}\|^{-1}\partial_\theta$ is FW transported (i.e., parallel transported) along the circular geodesic. With this observation, we can complete the orthonormal tetrad adding the spatial vector $\tilde{\mathbf{e}}_{r}=\|g_{rr}\|^{-1}\partial_r$ and the spatial vector $\tilde{\mathbf{e}}_{\phi}$ that can be obtained by requiring it to be orthonormal with the previous three. The tetrad thus formed is not FW transported. However, we can now linearly superpose $\tilde{\mathbf{e}}_{r}$ and $\tilde{\mathbf{e}}_{\phi}$ like $a(\phi)\tilde{\mathbf{e}}_{r}+b(\phi)\tilde{\mathbf{e}}_{\phi}$ and impose this vector to be FW transported. This results in two linearly independent differential equations for the coefficients $a(\phi),\,b(\phi)$ 
\begin{align*}
    &\sqrt{2} \sqrt{R} a(\phi ) \sqrt{2 R-3 r_S}+2 R b'(\phi )=0\\
    &\frac{2 a'(\phi )}{\sqrt{R}}-\frac{\sqrt{2} b(\phi ) \sqrt{2 R-3 r_S}}{R}=0,
\end{align*}
whose solution is readily obtained as 
\begin{align*}
    &a(\phi )=c_1 \cos \left(\frac{\phi  \sqrt{2 R-3 r_S}}{\sqrt{2} \sqrt{R}}\right)+\sqrt{1-c_1^2}\sin \left(\frac{\phi  \sqrt{2 R-3 r_S}}{\sqrt{2} \sqrt{R}}\right),\\
    &b(\phi )=\sqrt{1-c_1^2} \cos \left(\frac{\phi  \sqrt{2 R-3 r_S}}{\sqrt{2} \sqrt{R}}\right)-c_1 \sin \left(\frac{\phi  \sqrt{2 R-3 r_S}}{\sqrt{2} \sqrt{R}}\right),
\end{align*}
where also the normalization of the linear combination has been used and $c_1$ is an integration constant that we fix to one in one case and to zero in the other in order to obtain two new vectors $\mathbf{e}_{r}$ and $\mathbf{e}_{\phi}$ which complete the FW transported tetrad. Finally, the FW transported tetrad for the circular orbit geodesic observer in SdS spacetime in static coordinates and in matrix form where each line is one of the tetrad vectors, is given by
{\footnotesize
\begin{equation}
\mathbf{e}_J^\mu=\left(
\begin{array}{cccc}
 \sqrt{\frac{2R}{2 R-3 r_S}} & 0 & 0 & \sqrt{\frac{r_S/R-2 H_0^2 R^2}{R(2 R-3 r_S)}} \\
 -\sqrt{\frac{r_S-2 H_0^2 R^3}{(2 R-3 r_S)\alpha(R)}} \sin \left(\phi  \sqrt{1-\frac{3 r_S}{2R}}\right) & \sqrt{\alpha(R)}\cos \left(\phi  \sqrt{1-\frac{3 r_S}{2R}}\right) & 0 & -\sqrt{\frac{2\alpha(R)}{R(2R-3r_S)}}\sin \left(\phi  \sqrt{1-\frac{3 r_S}{2R}}\right) \\
 0 & 0 & \frac{1}{R} & 0 \\
 \sqrt{\frac{r_S-2 H_0^2 R^3}{(2 R-3 r_S) \alpha(R)}} \cos \left(\phi  \sqrt{1-\frac{3 r_S}{2R}}\right) & \sqrt{\alpha(R)}\sin \left(\phi  \sqrt{1-\frac{3 r_S}{2R}}\right) & 0 & \sqrt{\frac{2\alpha(R)}{R(2R-3r_S)}} \cos \left(\phi  \sqrt{1-\frac{3 r_S}{2R}}\right) \\
\end{array}
\right)
\end{equation}
}
It is important to note that a circular trajectory in SdS spacetime is characterized by an angular velocity $\omega^2=r_S/(2R^3)-H_0^2$~\cite{brito2020synchrotron}. This is exactly the quantity entering the trigonometric functions in the tetrad elements. Indeed $\dot{\phi}=L/R^2$ from the angular momentum conservation, which implies $\phi=\tau L/R^2$, where $\tau$ is the proper time of the observer. Then, from the expression for the angular momentum of the circular geodesics, we can easily check that the argument of the trigonometric functions appearing in the tetrad is $\omega\tau=\sqrt{ r_S/(2R^3)-H_0^2}\tau$.

\subsection{Bonus: Circular orbit geodesic observer in Kerr-deSitter spacetime}\label{B5}
A straightforward generalization of the SdS spacetime, which account for axially symmetric rotating central objects, instead of a spherical symmetric one, is the so called Kerr-de Sitter (KdS) spacetime~\cite{carter1968hamilton,demianski1973some,gibbons2005general}. The properties of the geodesic of KdS spacetime have been extensively considered in the existing literature and also the effect of the cosmological constant on the local dynamics have been considered~\cite{stuchlik2004equatorial,kraniotis2004precise,PhysRevD.81.044020,kerr2003standard}.

Following~\cite{PhysRevD.81.044020}, we can write the KdS metric in Boyer-Lindquist stationary coordinates as
\begin{equation}
    ds^2=-\frac{\Delta_r}{\chi^2\rho^2}(dt-a \sin^2\theta d\phi)^2+\frac{\rho^2}{\Delta_r}dr^2+\frac{\rho^2}{\Delta_\theta}d\theta^2+\frac{\Delta_\theta\sin^2\theta}{\chi^2\rho^2}\left(a dt-(r^2+a^2)d\phi\right)^2
\end{equation}
where 
\begin{align}
    & \Delta_r=(1-H^2 r^2)(r^2+a^2)-r_S r\\
    & \Delta_\theta=1+a^2H^2\cos^2\theta\\
    &\chi=1+a^2H^2\\
    &\rho^2=r^2+a^2\cos^2\theta
\end{align}
and $a=J/M$ is the angular momentum per mass of the central spinning object. Also recall that $H^2=\Lambda/3$, where $\Lambda$ is the cosmological constant. Note that the KdS spacetime is stationary but \textit{not} static, and it reduces to SdS spacetime in static coordinates for $a\to 0$.

As discussed in e.g.~\cite{kraniotis2004precise}, the geodesics of the KdS metric are characterized by four constants of integration given by energy per unit mass ($E$), angular momentum per unit mass ($L$), normalization of the tangent ($\mu$) and the modified Carter's constant ($Q$). Timelike equatorial geodesics are found by imposing $\theta=\pi/2$, $\mu=-1$, and $Q=0$. Further imposing the equatorial orbit to be circular fixes also the last two constants of integration.   

The FW tetrad for an equatorial circular orbit in KdS spacetime can be obtained by following the same steps as in the case of SdS spacetime. Due to the lengthy expressions for both the tetrad and the components of the Riemann tensor in the proper detector frame of a geodesic observer following a circular trajectory, we do not report them here. We limit ourselves to note that it is easy to verify that the Riemann tensor in the proper detector frame does not contain any linear term in $H$ and thus also the frequency shift for a resonator in KdS spacetime has no linear term in $H$.

\section{Doppler tracking and redshift: further details}\label{app:red}
In the main text, we have considered the redshift and DDT in FLRW spacetime for the cosmological and the Kodama observer field. In this appendix, we offer some further detail on those expressions and their derivation. 

In the following, given an observer field $\mathbf{u}$, the frequency of a light signal characterized by the vector $\mathbf{k}$ as measured by the observer is the scalar product between the observer field and $\mathbf{k}$, i.e. $\omega=g(\mathbf{u},\mathbf{k})$, where $g$ is the metric symmetric tensor.

\subsection{Redshift ratio in FLRW}
Let us consider the FLRW metric in isotropic, spherical coordinates
\begin{equation}
    ds^2=-dt^2+a^2(t)dr^2+a^2(t)r^2(\sin\theta^2 d\phi^2+d\theta^2).
\end{equation}
In these coordinates, the cosmological observer is given by $\mathbf{u}=\partial_t$ and it is a geodesic observer whose proper time coincides with the coordinate time. The expression for the redshift ratio for a light signal exchanged between two observers following the integral lines of $\mathbf{u}$ can be obtained in several ways, from using the fact that the cosmological observer is a conformal Killing 
vector field to brute force computations. For example, consider the two observers to be at $r_0$ and $r_1$ respectively. A radial null signal exchanged between these two is characterized by a null vector $\mathbf{k}=(1/a(t),1/a(t)^2,0,0)$ coming from the geodesic equation and the normalization of the null vector $g(\mathbf{k},\mathbf{k})=0$. Thus, one immediately finds
\begin{equation}
    \frac{\omega_1}{\omega_0}\equiv\frac{g(\mathbf{u},\mathbf{k})_{p_1}}{g(\mathbf{u},\mathbf{k})_{p_0}}=\frac{a(t_0)}{a(t_1)},
\end{equation}
where $p_i=(r_i,t_i),\,i=0,1$ are the spacetime points (suppressing the angular coordinates) at which the observers are located when receiving the light signal.

In the case of the Kodama observer, in isotropic coordinates this observer is expressed as 
\begin{equation}
    \mathbf{u}_{\rm K}=
    \frac{(1,-H r, 0, 0)}{\sqrt{1-H^2 R^2}},
\end{equation}
where $R=a(t)r$ and $\alpha(R)=1-H^2 R^2$ is the norm of the Kodama vector field\footnote{This is equal to $\alpha(R)=1-r_S/R-H_0^2 R^2$ in the SdS case.}. The redshift formula thus reads
\begin{align}\label{app:approx}
    \frac{\omega_1}{\omega_0}&\equiv\frac{g(\mathbf{u}_{\rm K},\mathbf{k})|_{p_1}}{g(\mathbf{u}_{\rm K},\mathbf{k})|_{p_0}}=\frac{\sqrt{1-H(t_0)^2 R_0^2}}{\sqrt{1-H(t_1)^2 R_1^2}}\left(\frac{1+H(t_1) R_1}{1+H(t_0) R_0}\right)\frac{a_0}{a_1}\\ \nonumber
    &\approx (1+\mathcal{O}(H^2,H')) \left( 1+H(t_0)R_1-H(t_0)R_0 + \mathcal{O}((HR)^2,H'R^2)\right) (1-H(t_0) \Delta t_{10}+\mathcal{O}((HR)^2,H'R^2))\\ \nonumber
    &\approx 1 + \mathcal{O}((HR)^2,H'R^2)\nonumber,
\end{align}
where in the second line we have expanded both the scale factor and the Hubble parameter and finally used the fact that\footnote{For a radial null geodesic $0 = ds^2= -dt^2+a^2(t) dr^2 = - [dt+dR/(1-RH)][dt-dR/(1+RH)]$. In lowest order (for constant $H(t)$) this implies $\Delta t_{10} =  \log [(1+H R_1)/(1+H R_0)]/H \approx \Delta R_{10} + \mathcal{O}(HR)$.} {$\Delta t_{10}=R_1-R_0+\mathcal{O}(HR)$}.
This expression shows that, for the radially propagating light rays, the linear contributions in $HR$ from the second and third term cancel. This result is in accordance with the simpler analytical derivation shown in the text, where we considered the special case of de Sitter. Indeed, for de Sitter, or SdS for that matter, the Kodama vector field is a timelike Killing vector field so that we can use the fact that the scalar product between a Killing field and $\mathbf{k}$ is constant along the null geodesic~\cite{wald2010general}. Thus, we have
\begin{align}
     \frac{\omega_1}{\omega_0}&=\frac{g(\mathbf{u}_{\rm K},\mathbf{k})|_{p_1}}{g(\mathbf{u}_{\rm K},\mathbf{k})|_{p_0}}=\frac{\sqrt{\alpha(R_0)}}{\sqrt{\alpha(R_1)}},
\end{align}
where in the final result we are left with only the ratio of the norms of the Kodama vector field thanks to the conservation law discussed, and $\alpha(R)=1-r_S/R-H_0^2R^2$ with $r_S=0$ in the case of de Sitter spacetime.

\subsection{Double Doppler Tracking}
Building on the results of the previous section, it is clear that when considering the Kodama observer in de Sitter (or SdS) spacetime, eq.~\eqref{ddt-Kod} of the main text can be obtained by combining the results coming from the relativistic Doppler effect (see Ref.~\cite{RevModPhys.82.169}) with the fact that 
\begin{equation}
    \frac{\omega_2}{\omega_1'}\frac{\omega_1}{\omega_0}=\frac{\sqrt{\alpha(R_0)}}{\sqrt{\alpha(R_2)}}=1,
\end{equation}
where we have used the fact that the Kodama observer is at a fixed value of the areal radius so that $R_0=R_2$. If considering the general case of FLRW spacetime, eq.~\eqref{app:approx} shows us that there would be a correction which, however, is at least quadratic in $H$, so that we can safely neglect it. 

This shows that no correction linear in $H$ should be expected in the Double Doppler Tracking (DDT) redshift ratio for light signals exchanged between Kodama observers. In order to prove that this is also the case for the rate of change of the DDT ratio, we need to show that  
differentiating the redshift ratio with respect to the proper time of the observer at the reception point does not introduce any linear correction. Note that this is not \textit{a priori} obvious since in the case of the cosmological observer, part of the linear corrections in $H$ are introduced exactly by this derivative.

Let us consider the Kodama observer in de Sitter spacetime (the derivation can be extended straightforwardly to the case of SdS). The Kodama observer field is given, in static coordinates, by $\mathbf{u}_{\rm K}=\alpha^{-1/2}\partial_t$. Thus, the coordinate time and the proper time of the Kodama observer are the same at linear order in $H_0$ and we can focus on the derivative with respect to the coordinate time of the receiver in the DDT scheme.

Note that the expression we are interested in  
differentiating with respect to $t_2$, i.e. eq.~\eqref{ddt-Kod}, contains only quantities that depend on $t_1$. Thus, we need to obtain $dt_1/dt_2$. Following~\cite{RevModPhys.82.169}, we can use the null condition for an inward directed (assuming $R_1>R_2=R_0$) radial, lightlike geodesic to get 
\begin{equation}
    \int_{t_1(t_2)}^{t_2}dt=-\int_{R_1(t_1(t_2))}^{R_2}\frac{dR}{\alpha(R)},
\end{equation}
and then take the derivative with respect to $t_2$, with the understanding that $R_2$ is constant for the Kodama observer. In this way we arrive at\footnote{{In the case of SdS we would have $\alpha(R_1)$ in the denominator of the right hand side of ~\eqref{c8}}.} 
\begin{align}\label{c8}
    \frac{dt_1}{dt_2}=\left(1+\frac{dR_1/dt_1}{1-H^2R_1^2}\right)^{-1}.
\end{align}
At this point, note that the spatial, unit vector $\hat{\bm{k}}$ -- i.e., the normalized spatial projection of the lightlike vector $\bm{k}=\left(E/\alpha,E,0,0\right)$ in the rest frame of the Kodama observer -- is $\hat{\bm{k}}=\sqrt{\alpha}\partial_R$
and this implies that the {projection of the relative spatial velocity of the spacecraft at the reflection point with respect to the observer field  $\bm{\beta}_{\mathbf{u}_{\rm K}}(\mathbf{v})=(\mathbf{v}-|g(\mathbf{v},\mathbf{u}_{\rm K})|\mathbf{u}_{\rm K})/|g(\mathbf{v},\mathbf{u}_{\rm K})|$ along $\hat{\bm{k}}$ has the form}
\begin{equation}
    \beta^{\hat{\bm{k}}}_{\mathbf{u}_{\rm K}}(\mathbf{v})\equiv-\frac{g(\hat{\bm{k}},\mathbf{v})}{g(\mathbf{u}_{\rm K},\mathbf{v})}=\frac{1}{\alpha(R_1)}\frac{dR_1}{dt_1},
\end{equation}
where $\mathbf{v}$ is the spacecraft 4-velocity.
With this last expression we have 
\begin{align}
    \frac{dt_1}{dt_2}=\left(1+\beta^{\hat{\bm{k}}}_{\mathbf{u}_{\rm K}}(\mathbf{v})\right)^{-1},
\end{align}
from which we see that this derivative does not introduce any correction linear in $H$. 

Finally, it should be noted that, in Ref.~\cite{RevModPhys.82.169} in order to arrive at the expression reported in eq.~\eqref{flrwDoubleDoppler} the authors need to  
differentiate the relative velocity with respect to $t_1$. In the case of the Kodama observer, we can follow the same derivation as in~\cite{RevModPhys.82.169,carrera2010geometrical} with the only caveat that additional corrections will appear when computing the derivatives of $\beta^{\hat{\bm{k}}}_{\mathbf{u}_{\rm K}}$ and $\beta^2_{\mathbf{u}_{\rm K}}=\|\bm{\beta}_{\mathbf{u}_{\rm K}}\|^2$ with respect to $t_1$. In particular, we would have that $d/dt_1=\left(dt_1/d\tau_1\right)^{-1}d/d\tau_1$, where $\tau_1$ is the arc-length of the spacecraft trajectory. Contrary to the case of the cosmological observer, this is not identical to the covariant ``observer'' derivative, $\nabla_{\mathbf{v}}^{\mathbf{u}}$ defined in eq.~(130) of Ref.~\cite{RevModPhys.82.169} (see also~\cite{carrera2006doppler,carrera2010geometrical}), anymore. Indeed, acting on a scalar function $f$
    \begin{equation}
        \nabla_{\mathbf{v}}^{\mathbf{u}_{\rm K}}f=|g(\mathbf{u}_{\rm K},\mathbf{v})|^{-1}\frac{df}{d\tau_1}=\frac{1}{\alpha^{1/2}}\frac{df}{dt_1},
    \end{equation}
from which,
\begin{equation}
    \frac{d}{dt_1}f=\alpha^{1/2}\nabla_{\mathbf{v}}^{\mathbf{u}_{\rm K}}f .
\end{equation}
Thus, in computing $d\beta^{\hat{\bm{k}}}_{\mathbf{u}_{\rm K}}(\mathbf{v})/dt_1$ and $d\beta^{2}_{\mathbf{u}_{\rm K}}(\mathbf{v})/dt_1$, we get:
\begin{itemize}
    \item $d\beta^{\hat{\bm{k}}}_{\mathbf{u}_{\rm K}}(\mathbf{v})/dt_1=\alpha^{1/2}\nabla_{\mathbf{v}}^{\mathbf{u}_{\rm K}}\beta^{\hat{\bm{k}}}_{\mathbf{u}_{\rm K}}(\mathbf{v})=\alpha^{1/2}\left(h_{\mathbf{u}_{\rm K}}(\bm{\alpha}_{\mathbf{u}_{\rm K}}(\gamma),\hat{\bm{k}})+h_{\mathbf{u}_{\rm K}}(\bm{\beta}_{\mathbf{u}_{\rm K}}(\mathbf{v}),\nabla_{\mathbf{v}}^{\mathbf{u}_{\rm K}}\hat{\bm{k}})\right)$, where $h$ is the spatial metric in the rest frame of the observer field, and we have introduced the relative spatial acceleration of the spacecraft trajectory with respect to the observer field $\bm{\alpha}_{\mathbf{u}_{\rm K}}(\gamma)$
    \item $d\beta^2_{\mathbf{u}_{\rm K}}(\mathbf{v})/dt_1=\alpha^{1/2}\nabla_{\mathbf{v}}^{\mathbf{u}_{\rm K}}\beta^2_{\mathbf{u}_{\rm K}}(\mathbf{v})=2\alpha^{1/2}h_{\mathbf{u}_{\rm K}}(\bm{\beta}_{\mathbf{u}_{\rm K}}(\mathbf{v}),\bm{\alpha}_{\mathbf{u}_{\rm K}}(\gamma))$,
\end{itemize}
where we have used the identities in eqs.~(3.16) and (3.17) in~\cite{carrera2010geometrical} {and, as in the main text, we indicate with $\bm{\alpha}_{\mathbf{u}_{\rm K}}(\gamma)= \nabla_{\mathbf{v}}^{\mathbf{u}_{\rm K}}\bm{\beta}_{\mathbf{u}_{\rm K}}(\mathbf{v})$ the relative spatial acceleration of the spacecraft trajectory $\gamma$ with respect to the observer field}.
However, it is clear that the corrections appearing will be at least quadratic in $H$ and cannot give rise to any linear correction in $H$ in the DDT expression. {Indeed, putting together the results listed above and taking the derivative of eq.~\eqref{ddt-Kod}, i.e.,  
\begin{equation}
    \frac{\omega_2}{\omega_0}=\left\{2\frac{1-\beta_{\mathbf{u}_{\rm K}}^{\hat{\bm{k}}}(\mathbf{v})}{1-\beta_{\mathbf{u}_{\rm K}}(\mathbf{v})^2}-1\right\},
\end{equation}
we arrive at
\begin{align}
    \frac{1}{\omega_0}\frac{d\omega_2}{d\tau_2}&=\alpha(R)^{-1/2}(1+\beta^{\hat{\bm{k}}}_{\mathbf{u}_{\rm K}}(\mathbf{v}))^{-1}\left[-2\left(h_{\mathbf{u}_{\rm K}}(\bm{\alpha}_{\mathbf{u}_{\rm K}}(\gamma),\hat{\bm{k}})+h_{\mathbf{u}_{\rm K}}(\bm{\beta}_{\mathbf{u}_{\rm K}}(\mathbf{v}),\nabla_{\mathbf{v}}^{\mathbf{u}_{\rm K}}\hat{\bm{k}})\right)(1-\beta^{2}_{\mathbf{u}_{\rm K}}(\mathbf{v}))^{-1}\right.\\ \nonumber
    &\left. +4h_{\mathbf{u}_{\rm K}}(\bm{\beta}_{\mathbf{u}_{\rm K}}(\mathbf{v}),\bm{\alpha}_{\mathbf{u}_{\rm K}}(\gamma))\frac{(1-\beta^{\hat{\bm{k}}}_{\mathbf{u}_{\rm K}}(\mathbf{v}))}{(1-\beta^{2}_{\mathbf{u}_{\rm K}}(\mathbf{v}))^2}\right]
\end{align}
to be compared with eq.(8.12) and eq.(144) of~\cite{carrera2010geometrical} and~\cite{RevModPhys.82.169} respectively. In particular, an even clearer picture can be obtained following~\cite{RevModPhys.82.169,carrera2010geometrical} and considering a radially escaping spacecraft so that $\bm{\beta}_{\mathbf{u}_{\rm K}}=\beta^{\hat{\bm{k}}}_{\mathbf{u}_{\rm K}}\hat{\bm{k}}$, $\bm{\alpha}_{\mathbf{u}_{\rm K}}=\alpha^{\hat{\bm{k}}}_{\mathbf{u}_{\rm K}}\hat{\bm{k}}$ with $\alpha^{\hat{\bm{k}}}_{\mathbf{u}_{\rm K}}=h_{\mathbf{u}_{\rm K}}(\bm{\alpha}_{\mathbf{u}_{\rm K}},\hat{\bm{k}})$ -- where we now suppress the arguments of the $\alpha^{\hat{\bm{k}}}_{\mathbf{u}_{\rm K}}$ and the $\beta^{\hat{\bm{k}}}_{\mathbf{u}_{\rm K}}$ for ease of notation. In this case we obtain
\begin{equation}
    \frac{1}{\omega_0}\frac{d\omega_2}{d\tau_2}=-2\alpha(R)^{-1/2}\alpha^{\hat{\bm{k}}}_{\mathbf{u}_{\rm K}}(1+\beta^{\hat{\bm{k}}}_{\mathbf{u}_{\rm K}})^{-3}= 2\left(1+\frac{H_0^2R^2}{2}\right)\alpha^{\hat{\bm{k}}}_{\mathbf{u}_{\rm K}}(1+\beta^{\hat{\bm{k}}}_{\mathbf{u}_{\rm K}})^{-3}+\mathcal{O}(H_0^4 R^4),
\end{equation}
which shows that only corrections quadratic, or higher, in $H_0R$ appear in the DDT rate.
}

\section{Weak field regime of McVittie spacetime}\label{app:perturbative}

In the following, We will diagonalize the McVittie metric in areal radius coordinates in eq.~\eqref{areal} to second order in $HR$. First, we find that the inverse of the McVittie metric is
\begin{equation}
    g^{\mu\nu}=\left(
\begin{array}{cccc}
 \frac{1}{1-\frac{2m_0}{R}} & -\frac{H(t)R}{\sqrt{1-\frac{2m_0}{R}}} & 0 & 0 \\
 -\frac{H(t)R}{\sqrt{1-\frac{2m_0}{R}}} & 1-\frac{2m_0}{R} - H(t)^2R^2 & 0 & 0 \\
 0 & 0 & \frac{1}{R^2} & 0 \\
 0 & 0 & 0 & \frac{1}{R^2\sin^2\theta} \\
\end{array}
\right).
\end{equation}
We define the time coordinate
\begin{equation}
    \tau = t-\int dR \,\frac{g^{01}}{g^{11}} \approx t - \Sigma(R) H(t) R^2/2
\end{equation}
which we approximated to second order in $HR$ and defined
\begin{equation}
    \Sigma(R) = 16\frac{m_0^2}{R^2}\left(1-\frac{2m_0}{R}\right)^{-1/2} {}_2F_1\left(-\frac{1}{2},3,\frac{1}{2},1-\frac{2m_0}{R}\right)\,,
\end{equation}
where ${}_2F_1$ is the hypergeometric function, 
and find that the inverse McVittie metric becomes
\begin{equation}
    g^{\mu\nu} \approx \left(
\begin{array}{cccc}
 -\frac{1-\Sigma(R)H'(t)R^2}{1-\frac{2m_0}{R}} - \frac{H(t)^2 R^2}{\left(1-\frac{2m_0}{R}\right)^2}   & 0 & 0 & 0 \\
 0 & 1-\frac{2m_0}{R} - H(t)^2R^2 & 0 & 0 \\
 0 & 0 & \frac{1}{R^2} & 0 \\
 0 & 0 & 0 & \frac{1}{R^2\sin^2\theta} \\
\end{array}
\right)
\end{equation}
also in second order in $HR$. The corresponding approximate expression for the McVittie metric is
\begin{equation}
    g_{\mu\nu} \approx \left(
\begin{array}{cccc}
 -\left(1 - \frac{2m_0}{R}\right)(1 + \Sigma(R)H'(t)R^2) + H(t)^2 R^2 & 0 & 0 & 0 \\
 0 & \frac{1}{1-\frac{2m_0}{R}} + \frac{H(t)^2 R^2}{\left(1-\frac{2m_0}{R}\right)^2} & 0 & 0 \\
 0 & 0 & R^2 & 0 \\
 0 & 0 & 0 & R^2\sin^2\theta\\
\end{array}
\right).
\end{equation}
In the weak-field regime, where $m_0/R\ll 1$, we can consider this as a linear perturbed Minkowski metric
and identify a Newtonian potential via the relation $g_{00} = -1 - 2\Phi$. We find
\begin{equation}\label{eq:McVittiediag}
    g \approx \rm{diag}\left(
 -1 + \frac{2m_0}{R} + (H(t)^2 + H'(t)) R^2 ,
 1 + \frac{2m_0}{R} + H(t)^2R^2 , \frac{1}{R^2} ,   \frac{1}{R^2\sin^2\theta} \right).
\end{equation}
where we neglected terms proportional to $m_0 H(t)^2$ and $m_0 H'(r)$. The Newtonian potential becomes
\begin{equation}
    \Phi = - \frac{m_0}{R} - \frac{1}{2} (H(t)^2 + H'(t)) R^2\,.
\end{equation}
Note the similarity of \eqref{eq:McVittiediag} to the SdS metric in manifestly static coordinates in \eqref{SdS}. In contrast to the latter case, in the case of the perturbatively diagonalized McVittie metric, there is also a time-dependent spatial component.

\end{document}